\documentclass[a4paper]{article}

\usepackage{amsmath,amsfonts,amssymb,amsthm}

\newtheorem{teo}{Theorem.}
\newtheorem{prop}[teo]{Proposition.}
\newtheorem{cor}[teo]{Corollary.}
\newtheorem{lem}[teo]{Lemma.}
\newtheorem{rmk}{Remark.}

\newcommand{\Op}[1]{\mathcal{#1}}

\begin{document}
\title{Polynomial constants of motion for Calogero-type systems in three dimensions}
\author{Claudia Chanu \\ \\
Dipartimento di Matematica e Applicazioni,\\ Universit\`a di Milano Bicocca.  Milano, via Cozzi 53, Italia.
\\
\\
 Luca Degiovanni \, Giovanni Rastelli \\ \\ Last affiliation: Dipartimento di Matematica, \\ Universit\`a di Torino.  Torino, via Carlo Alberto 10, Italia.\\ \\ e-mail: claudia.chanu@unimib.it \\ luca.degiovanni@gmail.com \\ giorast.giorast@alice.it }
\maketitle

\begin{abstract}
We give an explicit and concise formula for higher-degree polynomial first integrals of a family of Calogero-type Hamiltonian systems in dimension three. These first integrals, together with the already known ones, prove the maximal superintegrability of the systems.
\end{abstract}
\section{Introduction}
In the paper \cite{3b_line} the system with Hamiltonian
\begin{equation}\label{eq1}
H_0=\frac{1}{2}\left(p_r^2+\frac{1}{r^2}p^2_\psi+p^2_z\right)+\frac{F(\psi)}{r^2}
\end{equation}
is studied. This family of systems include many well-known integrable systems with three degrees of freedom, as well as the rational Calogero system, the Wolfes system and the Tremblay-Turbiner-Winternitz system without the harmonic oscillator term. The Hamiltonian (\ref{eq1}) admits the following four constants of motion 
\begin{eqnarray*}
H_1&=&\frac{1}{2}p_z^2, \qquad H_2=\frac{1}{2}p_\psi^2+F(\psi),\\
H_3&=&\frac{1}{2}\left[(rp_z-zp_r)^2+\left(1+\frac{z^2}{r^2}\right)p_{\psi}^2\right] +\left(1+\frac{z^2}{r^2}\right)F(\psi),\\
H_4&=&\frac{1}{2} \left(zp_r^2 +\frac{z}{r^2}p_{\psi}^2-rp_rp_z\right) + \frac{z}{r^2}F(\psi),
\end{eqnarray*}
but, since the rank of the Jacobian of $(H_0,\ldots,H_4)$ is only four, these integrals establish only the quasi-maximal superintegrability of the system. In order to prove the maximal superintegrability, a further integral is needed.

It is clear that the maximal superintegrability of the Hamiltonian (\ref{eq1}) depends only on the maximal superintegrablity of the two-dimensional Hamiltonian
$$ 
H=H_0-H_1=\frac{1}{2}\left(p_r^2+\frac{1}{r^2}p^2_\psi\right)+\frac{F(\psi)}{r^2}
$$
i.e., on the existence of an extra constant of the motion besides the angular component
$$
H_2=\frac{1}{2}p^2_\psi+F(\psi)\,.
$$
In the preprint \cite{3b_line_old} (first version) an extra independent integral was found by direct computation for the particular case
$$
F(\psi)=\frac{k_1}{\cos^2\psi}+\frac{k_2}{\sin^2\psi}
$$
and for the pairs of equivalent potentials 
\begin{equation}\label{CW}
F(\psi)=\frac{k}{\cos^2\lambda\psi} \qquad\mbox{and}\qquad F(\psi)=\frac{k}{\sin^2\lambda\psi}
\end{equation}
in the cases $\lambda=2,3,4,5,\frac{1}{2},\frac{1}{3},\frac{1}{5}$. Moreover in \cite{3b_line} was conjectured that the general expression for the case 
$$\label{QQ}
F(\psi)=\frac{k}{\sin^2\lambda\psi}, \qquad \lambda=2n+1,
$$
is given, up to $(-1)^n$, by the quite complicated explicit formula (here slightly simplified)
\begin{equation}\label{formulaccia}
\sum_{\sigma =0}^{n}\sum_{i=0}^{2\sigma+1} 
\frac{p_r^i p_\psi^{l-i}}{r^{\lambda-i}}(-2F)^{n-\sigma} 
{\lambda\choose i} {[(\lambda-i)/2]\choose [(l-i)/2]}
\left(\frac{1}{\lambda}\;\frac{d}{d\psi}\right)^{l-i}\cos\lambda\psi
\end{equation}
where $l=2\sigma+1$, $\left(^a _b\right) =\frac {a!}{b!(a-b)!}$ denotes the Newton binomial symbol and $[a]$ the integer part of $a$.
The expression (\ref{formulaccia}) is indeed a first integral of (\ref{eq1}) functionally independent from  $H_0$, $H_1$, $H_2$ and  $H_4$. The proof can be done by expanding the Poisson brackets of  (\ref{formulaccia}) with (\ref{eq1}) and verifying that all coefficients of each monomial in $p_r$ and $p_\psi$ are identically zero. The functional independence is proved in the same way as for the expression (\ref{ip}) below. 

The system of Hamiltonian
\begin{equation}\label{KM}
p_r^2+\frac 1{r^2}p_\psi^2+\alpha r^2+\frac \beta{r^2\cos ^2 k\psi}+ \frac \gamma{r^2\sin ^2 k\psi},
\end{equation}
with $\alpha$, $\beta$, $\gamma $ real and $k$ rational, that includes  (\ref{QQ}) as a subcase, has been intensively studied in \cite{W1} and \cite{W2} and proved to be maximally superintegrable in \cite{KM1}. By a different approach, in the following we provide a compact expression for the additional functionally independent polynomial first integral in the cases $\alpha=\beta=0$ and $k$ integer, improving and generalizing to even integers the result obtained with the formula (\ref{formulaccia}).

\section{A simple formula for the extra integral}
We consider the Hamiltonian
\begin{equation}\label{ham}
H=\frac{1}{2}p_r^2+\frac{1}{r^2}\left(\frac{1}{2}p^2_\psi+F(\psi)\right).
\end{equation}

The Hamiltonian vector field associated with $H$ is
\begin{equation}\label{ham_vet}
\Op{X}_H=p_r \frac{\partial}{\partial r}+\frac{2L}{r^3}\frac{\partial}{\partial p_r}+\frac{1}{r^2}\Op{X}_L,
\end{equation}
where $L$ is the angular part of $H$
\begin{equation}\label{ang}
L=\frac{1}{2}p^2_\psi+F(\psi)
\end{equation}
and $\Op{X}_L$ is the associated Hamiltonian vector field
\begin{equation}\label{ang_vet}
\Op{X}_L=p_\psi\frac{\partial}{\partial \psi}-\dot{F}\frac{\partial}{\partial p_\psi}\,.
\end{equation}


After recalling that, for any two differential operators $\Op{A}$ and $\Op{B}$, if their commutator $[\Op{A},\Op{B}]$ commutes with $\Op{B}$ then
\begin{equation}\label{eq3}
\Op{A}\Op{B}^\lambda=\lambda[\Op{A},\Op{B}]\Op{B}^{\lambda-1}+\Op{B}^\lambda\Op{A}\,=\lambda\Op{B}^{\lambda-1}[\Op{A},\Op{B}]+\Op{B}^\lambda\Op{A},
\end{equation}
we show that this property can be applied to $\Op{X}_H$ and to the operator 
\begin{equation}
\Op{U}=p_r+\frac{\mu}r \Op{X}_L, \qquad (\mu \in \mathbb{R}).
\end{equation}

\begin{lem}\label{lem1}
We have
\begin{eqnarray*}
\left[\Op{X}_L,L\right] &=& 0, \\
\left[\frac{2L}{r^3}\frac{\partial}{\partial p_r},\Op{U}\right] &=& \frac{2L}{r^3}, \\
\left[p_r\frac{\partial}{\partial r},\Op{U}\right] &=& -\frac{\mu}{r^2}p_r\Op{X}_L.
\end{eqnarray*}
Therefore,  $[[\Op{X}_H, \Op{U}],\Op{U}]=0$.
\end{lem}
\textbf{Proof.}
Since $\Op{X}_L$ is the Hamiltonian vector field associated to $L$, the identity $[\Op{X}_L,L]=\{L,L\}=0$ follows, where $\{\cdot,\cdot\}$ is the standard Poisson bracket. The last two commutators are obtained by straightforward evaluations.
Hence, 
$$
[[\Op{X}_H, \Op{U}],\Op{U}]=\left[\frac{2L}{r^3}-\frac{\mu}{r^2}p_r\Op{X}_L,p_r+\frac{\mu}r \Op{X}_L\right]=0.
$$
\qed

\begin{lem}\label{lem2}
The square of the differential operator $\Op{X}_L$ is
$$
\Op{X}_L^2=p^2_\psi\frac{\partial^2}{\partial\psi^2}-\dot{F}\frac{\partial}{\partial\psi}+
\dot{F}^2\frac{\partial^2}{\partial p_\psi^2}-p_\psi\left(\ddot{F}+2\dot{F}\frac{\partial}{\partial\psi}\right)\frac{\partial}{\partial p_\psi}
$$
and when applied to a function of $\psi$ only it coincides with the operator
$$
p^2_\psi\frac{\partial^2}{\partial\psi^2}-\dot{F}\frac{\partial}{\partial\psi}\,.
$$  
\end{lem}
\textbf{Proof.}
It follows from a straightforward calculation. \qed

\begin{teo}\label{teo1}
The function
\begin{equation}\label{ip}
I_\lambda=\Op{U}^\lambda G(\psi)=\left(p_r+\frac{\mu}{r}\Op{X}_L\right)^\lambda G(\psi)
\end{equation}
is a constant of the motion of the Hamiltonian (\ref{ham}),
$$
H=\frac{1}{2}p_r^2+\frac{1}{r^2}\left(\frac{1}{2}p^2_\psi+F(\psi)\right),
$$
if and only if the following conditions are satisfied
\begin{equation} \label{FG}
\lambda\mu=1\,, \qquad
F(\psi)=\frac{k}{\sin^2(\lambda\psi+\psi_0)}\,, \qquad
G(\psi)=\cos(\lambda\psi+\psi_0)\,. 
\end{equation}
\end{teo}

\textbf{Proof.}

Let $G(\psi)$ and $F(\psi)$ be arbitrary functions. By Lemma \ref{lem1} we can apply formula~(\ref{eq3}) to $\Op{X}_H$ and $\Op{U}$, and by Lemma \ref{lem2} we get
\begin{eqnarray*}
\Op{X}_H\Op{U}^\lambda &=&
\lambda\Op{U}^{\lambda-1}[\Op{X}_H,\Op{U}]+\Op{U}^\lambda\Op{X}_H \\
&=&
\lambda\Op{U}^{\lambda-1}\left[p_r\partial_r+\frac{2L}{r^3}\partial_{p_r}+\frac{1}{r^2}\Op{X}_L,p_r+\frac{\mu}{r}\Op{X}_L\right]
+\Op{U}^\lambda\Op{X}_H \\
&=& \lambda\Op{U}^{\lambda-1}\left(-\frac{\mu}{r^2}p_r\Op{X}_L+\frac{2L}{r^3}\right)+
\frac{1}{r^2}\Op{U}^\lambda\Op{X}_L+\Op{U}^\lambda\left(p_r\partial_r+\frac{2L}{r^3}\partial_{p_r}\right) \\
&=& \Op{U}^{\lambda-1}\left(-\frac{\lambda\mu}{r^2}p_r\Op{X}_L+\frac{\lambda}{r^3}(p_\psi^2+2F)+\frac{1}{r^2}p_r\Op{X}_L+\frac{\mu}{r^3}\Op{X}_L^2\right)+\\
&&\Op{U}^\lambda\left(p_r\partial_r+\frac{2L}{r^3}\partial_{p_r}\right) \\
&=&\Op{U}^{\lambda-1}\left(\frac{1-\lambda\mu}{r^2}p_r\Op{X}_L+\frac{1}{r^3}(\lambda+\mu\partial^2_\psi)p_\psi^2+\frac{1}{r^3}(2\lambda F-\mu\dot{F}\partial_\psi)\right)+\\
&& \frac \mu{r^3}\Op{U}^{\lambda-1}\left(\dot{F}^2\partial^2_{p_\psi}-p_\psi(\ddot{F}+2\dot{F}\partial_\psi)\partial_{p_\psi}\right)
+\Op{U}^\lambda\left(p_r\partial_r+\frac{2L}{r^3}\partial_{p_r}\right).
\end{eqnarray*}
The evaluation of the operator $\Op{X}_H\Op{U}^\lambda$ on the function $G(\psi)$ results in the polynomial in the momenta function
\begin{equation}\label {12bis}
\Op{U}^{\lambda-1}\left(\frac{1-\lambda\mu}{r^2}p_r\Op{X}_L+\frac{1}{r^3}(\lambda+\mu\partial^2_\psi)p_\psi^2+\frac{1}{r^3}(2\lambda F-\mu\dot F\partial_\psi)\right)G.
\end{equation}
The operator $\Op{U}$, acting on a polynomial function of the momenta, increases by one the degree of the latter in $p_r$. Therefore, the expression (\ref{12bis}) vanishes if and only if
$$
\frac {1-\lambda \mu}{r^2}\dot Gp_rp_\psi+\frac 1{r^3}(\lambda G+\mu \ddot G)p_\psi^2+\frac 1{r^3}(2\lambda FG-\mu\dot F\dot G)=0,
$$
that is, when $\dot G\neq 0$, if and only if the three following equations are satisfied
\begin{equation}
\left\{ \label{sis}
\begin{array}{l}
\lambda\mu=1,\\
\mu\ddot{G}+\lambda G=0,\\
2\lambda FG-\mu\dot F \dot G=0.
\end{array}
\right.
\end{equation}
The general solution of $\ddot{G}+\lambda^2 G=0$ is clearly, up to a multiplicative constant, $G(\psi)=\cos(\lambda\psi+\psi_0)$ and substituting in the last equation we find the form of $F(\psi)$ given in the thesis. 
The case $\dot G=0$ leads to $G=const.$ and  either $\lambda =0$, $G=0$ or $F=0$, conditions all corresponding to trivial solutions.  
\qed

\begin{rmk}
The constant of motion (\ref{ip}) is well defined in the Euclidean space only when $\lambda$ is  a positive integer. Theorem \ref{teo1} states that the Hamiltonian $H$ (\ref{ham}) is essentially the only one (up to a phase shift) that admits a constant of motion of the form (\ref{ip}). In particular a constant of motion for rational values of $\lambda$ takes necessarily a different form. Although there are strong evidences that a constant of motion for rational values of $\lambda$ can be obtained in a similar way, the exact form of the generating differential operator still remains an open problem.
\end{rmk}
\begin{rmk}
Making the change of coordinates $\phi=\lambda\psi$, $p_\phi=\lambda p_\psi$ and setting $k=\lambda^2\tilde{k}$, the Hamiltonian $H$ (\ref{ham}), with $F$ given by (\ref{FG}), takes the form
$$
\frac{1}{2}p_r^2+\frac{\lambda^2}{r^2}\left(\frac{1}{2}p^2_\phi+\frac{\tilde{k}}{\sin^2\phi}\right)
$$
in which the parameter $\lambda$ take the role of a coupling constant. Obviously, all the proprieties of $H$ hold  for this Hamiltonian also.  
\end{rmk}

\begin{cor} Let $F(\psi)$ and $G(\psi)$  be defined as in (\ref{FG}).
\begin{enumerate}
\item The function
$$
{\Op{X}_L}^\nu \left(p_r+\frac{1}{\lambda r}\Op{X}_L\right)^\lambda G(\psi)
$$
is a constant of motion for the Hamiltonian (\ref{ham}),
$$
H=\frac{1}{2}p_r^2+\frac{1}{r^2}\left(\frac{1}{2}p^2_\psi+F(\psi)\right),
$$
for any $\nu\in\mathbb{N}$.
\item For any $\lambda \in \mathbb N$, the function
\begin{equation}\label{geo_int}
\left(p_r+\frac{p_\psi}{\lambda r}\,\frac{\partial}{\partial\psi}\right)^\lambda G(\psi)
\end{equation}
is a constant of motion for the geodesic part
$$
H_g=\frac{1}{2}\left(p_r^2+\frac{p^2_\psi}{r^2}\right)
$$
of the Hamiltonian $H$.
\end{enumerate}

\end{cor}
\textbf{Proof.} i) It follows immediately from Theorem \ref{teo1} and from the fact that the operator $\Op{X}_L$ commute with the operator $\Op{X}_H$, as a consequence of Lemma \ref{lem1}.

ii) It is well-known that, given a polynomial constant of motion for a natural Hamiltonian $H_g+V$, the part of highest degree in the momenta is a constant of motion for $H_g$. The function (\ref{geo_int}) is clearly the part of highest degree in the momenta of the function
$$
\left(p_r+\frac{1}{\lambda r}(p_\psi\partial_\psi-\dot{F}\partial_{p_\psi})\right)^\lambda G(\psi),
$$
given by Theorem \ref{teo1}.
\qed


The following theorem states the maximal superintegrability of the Hamiltonian $H$ (\ref{ham}) with $F$ given as in (\ref{FG}) and, as a consequence, of $H_0$ (\ref{eq1}) under the same hypothesis.

\begin{teo}
The three functions $H$, $L$, $I_\lambda$ -- under the  assumptions (\ref{FG}) -- are functionally independent and hence the five functions $H_0,\ldots, H_3$ and $I_\lambda$ are also functionally independent. 
\end{teo}

\textbf{Proof.}
We give a direct check of the independence of the three integrals of motion $H$, $L$ and $I_\lambda$. Their $3\times 4$ Jacobian matrix w.r.t. the variables $(p_r,p_\psi,r,\psi)$ is
$$
\left[
\begin{array}{cccc}
p_r & r^{-2}p_\psi & -2r^{-3}L & r^{-2}\dot{F} \\
0 & p_\psi & 0 & \dot{F} \\
\partial_{p_r}I_\lambda & \partial_{p_\psi}I_\lambda & \partial_{r}I_\lambda & \partial_{\psi}I_\lambda
\end{array}
\right],
$$
the minor obtained by deleting the fourth column is
$$
p_\psi(p_r \partial_r I_\lambda +\frac{2L}{r^3}\partial_{p_r}I_\lambda),
$$
which is a polynomial in the momenta.
Since $I_\lambda$ is a polynomial of degree $\lambda$ in  $p_r$,
the highest term in $p_r$ in the second addendum in the brackets is of degree $\lambda-1$, while 
$$
p_r \partial_r I_\lambda= p_r\partial_r\left(\sum_{i=0}^\lambda 
{\lambda \choose i}
p_r^{\lambda-i} \left(\frac 1{\lambda r}\Op{X}_L\right)^i\right)G
$$
has not a $(\lambda +1)$th-degree term in $p_r$ (being the coefficient of $p_r^\lambda$ in $I_\lambda$ independent of r), but it contains a nonzero 
 term of degree $\lambda$
$$
p_r\lambda p_r^{\lambda-1} \frac 1{\lambda r}\Op{X}_L(G)=p_r^\lambda p_\psi \dot{G}.
$$
Hence, being the minor not identically zero, the functions are functionally independent (up to a closed singular set). The functional independence of the five polynomials $H_0,\ldots, H_3$ and $I_\lambda$, with $F$ and $G$ given by (\ref{FG}), follows immediately
\qed

\begin{rmk}
A comparison between (\ref{ip}) and the old result (\ref{formulaccia}) has been done for the moment only through the explicit computation of the two expressions, in a large number of cases. The two expressions always coincide. Hence the constant of motion given by (\ref{ip}) reasonably corresponds with the conjectured one. Recently, different expressions have been obtained for higher-order polynomial first integrals of the Hamiltonian of Theorem \ref{teo1}. In these expressions, the degree of the first integral, while depending on $\lambda$, is greater than $\lambda$ itself, see for example \cite{KM1},\cite{MPY}.  
\end{rmk}

 The existence of the $\lambda$th-order first integral (\ref{ip}) for $F=\frac k{\sin^2 \lambda\psi}$ can be understood in connection with the $2\lambda$th-order dihedral symmetry of the function $F$: hexagonal for $\lambda=3$ (i.e., the Calogero case), octagonal for $\lambda=4$ and so on. 

The potentials (\ref{KM}) with rational coefficients $k=\frac pq$, represented in polar coordinates on the Euclidean plane, if $\beta\neq \gamma$, have period $q\pi$, if $q$ is even, or $2q\pi$ otherwise. Therefore, they are not single-valued for $q>2$ but still show, at least formally, dihedral symmetry of order $p$ or $2p$ respectively.  If $\beta=\gamma$, dihedral symmetries and periods are the same as above with $2k$ instead of $k$. Indeed, by substituting $2\cos^2 k\psi =1+\cos 2k\psi $ and $2\sin^2 h\psi=1-\cos 2k\psi $ in (\ref{KM}), one obtains for the angular part of the potential 
$$
2\frac {(\beta + \gamma)+(\gamma-\beta)\cos 2k\psi}{\sin ^2 2k\psi},
$$
the function has the symmetries and the period of $\cos 2k\psi$, the same as $\sin ^2 k\psi$, unless $\beta=\gamma$ when the numerator is a constant and the denominator only determines period and symmetries.

For $\alpha=0$, the Hamiltonian (\ref{KM}) has been studied in \cite{MPY} and a necessary condition is obtained for its maximal superintegrability, namely, that $k$ must be rational. It follows that for non-rational $k$, (\ref{KM}) is not maximally superintegrable. For non-rational values of $k$, (\ref{KM}) in $\mathbb E^2$ becomes aperiodic, infinitely-many valued and loses the dihedral symmetry, confirming the connection between the existence of an extra first-integral and the invariance under dihedral symmetry groups.

It is an open problem if other systems with dihedral symmetry admit corresponding higher-order first integrals, and under what conditions. Some attempts have been done to the analysis of integrability and superintegrability in connection with discrete symmetries in three-dimensional manifolds \cite{PS}, \cite{KM2}.

While the Hamiltonian (\ref{ham}), with $F(\psi)$ given by (\ref{FG}), admits a dihedral symmetry group of order $2\lambda$, the corresponding first integral $I_\lambda$ (\ref{ip})  admits instead a dihedral symmetry group of order $\lambda$. Indeed,

\begin{prop} Let $\psi_h=\psi+\frac h\lambda \pi$ with $h$ integer. Then,
$$
I_\lambda(\psi_h)=(-1)^hI_\lambda(\psi).
$$
\end{prop}

\textbf{Proof.} The transformation clearly leaves unchanged both the Hamiltonian and the operator $\Op{U}$, while $\cos (\lambda \psi_h)=(-1)^h\cos (\lambda\psi)$. The statement follows immediately from Theorem \ref{teo1}. \qed

\begin{rmk} 
In \cite{3b_line} and \cite{CDR1} the superintegrability of three and $n$-body systems on a line and on $m$-dimensional manifolds is deduced from the superintegrability of one-particle systems in three and $mn$-dimensional Euclidean spaces. By this approach, equivalence classes of $n$-body systems are determined by finite rotations of the same one-particle system in the $mn$-dimensional space, in such a way that all equivalent $n$-body systems are described by the same system of differential equations in $\mathbb E_{mn}$ in which angular phases only change. It is a $\pi/6$ phase which realizes the equivalence between the Calogero and the Wolfes systems (\ref{CW}).
\end{rmk} 

\section{Conclusion and open problems}

We give an explicit and compact expression for the $\lambda$th-degree polynomial in the momenta first integral of (\ref{ham}) with
$$
F=\frac{k}{\sin^2 \lambda\psi},
$$
for each positive integer $\lambda $. This polynomial, together with other four already known quadratic in the momenta first integrals, makes the Hamiltonian system (\ref{eq1}) maximally superintegrable. The system considered here is a particular case of the more general natural Hamiltonian whose  potential is given by (\ref{KM}) that has recently proven to be maximally superintegrable not only for integer but also for rational values of the parameter $k$. It seems not impossible to generalize the formula (\ref{ip}) to include the whole potential (\ref{KM}). The problems of finding such a concise expression also for rationals parameters and of explicitly proving the correspondence between (\ref{formulaccia}) and (\ref{ip}) remain open.

\end{document}